\documentclass[journal,10pt]{IEEEtran}
\usepackage[pdftex]{graphicx}
\usepackage[caption=false,font=footnotesize]{subfig}
\usepackage{ifpdf}
\usepackage{cite}
\usepackage{amsmath}
\usepackage{mdwmath}
\usepackage{mdwtab}
\usepackage{array}
\usepackage{stfloats}
\usepackage{url}
\usepackage{color,xcolor}
\usepackage{siunitx}
\usepackage{footnote}
\usepackage{booktabs} 
\usepackage{threeparttable}
\usepackage{multirow}

\begin{document}
%
\title{Physics-informed Deep Learning for Musculoskeletal Modelling: Predicting Muscle Forces and Joint Kinematics from Surface EMG}
%
%

\author{Jie Zhang,~\IEEEmembership{Member, IEEE},
        Yihui Zhao,
        Fergus Shone,
        Zhenhong Li,~\IEEEmembership{Member, IEEE},
        Alejandro F. Frangi,~\IEEEmembership{Fellow, IEEE},
        Shengquan Xie,~\IEEEmembership{Senior Member, IEEE},
        Zhiqiang Zhang,~\IEEEmembership{Member, IEEE}

\thanks{Jie Zhang, Yihui Zhao, Zhenhong Li, Shengquan Xie, and Zhiqiang Zhang are with the School of Electronic and Electrical Engineering, University of Leeds, Leeds, LS2 9JT, U.K. (e-mails: \{eenjz, eenyzhao, z.h.li, s.q.xie, z.zhang3\}@leeds.ac.uk).}
\thanks{Fergus Shone is with the School of Computing, University of Leeds, Leeds, LS2 9JT, U.K. (e-mail: mm16f2s@leeds.ac.uk).}
\thanks{Alejandro F. Frangi is with the School of Computing, University of Leeds, Leeds, LS2 9JT, U.K., also with the Alan Turing Institute, London, NW1 2DB, U.K., and also with the Department of Electrical Engineering, KU Leuven, 3000 Leuven, Belgium (e-mail: a.frangi@leeds.ac.uk).}
}

\maketitle

\begin{abstract}
Musculoskeletal models have been widely used for detailed biomechanical analysis to characterise various functional impairments given their ability to estimate movement variables (i.e., muscle forces and joint moment) which cannot be readily measured in \textit{vivo}. Physics-based computational neuromusculoskeletal models can interpret the dynamic interaction between neural drive to muscles, muscle dynamics, body and joint kinematics and kinetics. Still, such set of solutions suffers from slowness, especially for the complex models, hindering the utility in real-time applications. In recent years, data-driven methods has emerged as a promising alternative due to the benefits in speedy and simple implementation, but they cannot reflect the underlying neuromechanical processes. This paper proposes a physics-informed deep learning framework for musculoskeletal modelling, where physics-based domain knowledge is brought into the data-driven model as soft constraints to penalise/regularise the data-driven model. We  use the synchronous muscle forces and joint kinematics prediction from surface electromyogram (sEMG) as the exemplar to illustrate the proposed framework. Convolutional neural network (CNN) is employed as the deep neural network to implement the proposed framework. At the same time, the physics law between muscle forces and joint kinematics is used the soft constraint.  Experimental validations on two groups of data, including one benchmark dataset and one self-collected dataset from six healthy subjects, are performed. The experimental results demonstrate the effectiveness and robustness of the proposed framework.
\end{abstract}

\begin{IEEEkeywords}
Musculoskeletal modelling, deep neural network, physics-informed domain knowledge, muscle forces and joint kinematics prediction.
\end{IEEEkeywords}
%
\IEEEpeerreviewmaketitle

\section{Introduction}
\IEEEPARstart{H}{uman} movements encompass complex interactions of the neuromuscular system~\cite{rajagopal2016full}. As a powerful omputational simulation tool, musculoskeletal model can be applied for detailed biomechanical analysis to understand these interactions, which would be beneficial to various applications ranging from evaluating rehabilitation treatment~\cite{berning2021myoelectric},
enhancing performance of athlete~\cite{zaman2021hybrid,mcerlain2021review}, optimising robotic design for impaired individuals~\cite{akbas2019neuromusculoskeletal}, to surgical planning and intervention~\cite{persad2021vivo}.

Thus far, the majority of the musculoskeletal models are based on physics-based modelling technique to interpret transformation among neural excitation, muscle dynamics, and joint kinematics and kinetics~\cite{zhang2022personalized,jung2021intramuscular}. Via employing experimental recordings, e.g., electromyograms (EMGs), foot-ground reaction forces (GRFs), and segmental body kinematics, these models can provide a non-invasive means to estimate physiological quantities, such as muscle forces and the joint moment~\cite{bennett2022emg}. However, these models often suffer from the redundancy issue since the countless number of potential neural solutions can be employed to execute a single movement. Thus, static optimisation is commonly applied to solve this redundancy problem, which involves the use of inverse dynamics to track external joint moments and/or joint kinematics and estimation of muscle forces to satisfy pre-selected objective criteria, such as minimisation of the muscle activation squared~\cite{chambers2021model,park2022direct}. An alternative approach is to use EMG-driven neuromusculoskeletal models, consisting of a neural-driven forward dynamics model and static optimisation element~\cite{zhao2022musculoskeletal,zhao2020emg}. EMG can be used to calibrate musculotendon parameters of the model to individual properties (i.e., tendon slack length and optimal fiber length, etc.), via the optimisation procedure to best match experimental and estimated joint moments. It also enhances the joint torque estimation with static optimisation by adjusting the experimental EMG signals and synthesising the muscle excitations~\cite{sartori2014hybrid}. Although EMG-driven models overcome the limitations of static optimisation and are readily available in the past years, they are not without shortcomings.
The enhanced analysis required from EMG-driven models is time-consuming and slow, especially for complex models in high-dimensional spaces, which thus limits the utility of modelling for real-time applications.

To address the time-consuming issues of physics-based musculoskeletal models, data-driven models have also been explored to establish relationships between movement variables and neuromuscular status, i.e., from EMGs to joint kinematics and muscle forces~\cite{hajian2022deep,wu2017grip, mcdonald2020myoelectric,bao2022towards}. A major advantage of these data-driven models over physics-based models is speed. Although training may be lengthy, as inference involves a relatively simple forward pass through the network, it is computationally inexpensive and thus very quick. For instance,
Hu~\textit{et al.}~\cite{hu2022novel} utilised the long short-term memory (LSTM) network to estimate grasping forces from high-density surface EMGs (sEMGs). Geng~\textit{et al.} ~\cite{geng2022cnn} proposed a convolution with attention mechanism network (CNN-Attention) for continuous finger kinematics prediction from sEMGs.
Rane~\textit{et al.} ~\cite{rane2019deep} employed a deep neural network to learn the feature mapping from movement
space to muscle space, so musculoskeletal force could be predicted from kinematics. Similar ideas were also reported in~\cite{bao2021inter,huang2019real, ameri2018real,su2021deep}.
However, data-driven models are established without explicit physics modelling of the underlying neuromechanical processes, and they are essentially ``black-box" tools where all intermediate functional relationships cannot reflect the mechanisms underlying the observed variables~\cite{karniadakis2021physics,raissi2019physics}.

To address the drawbacks above of both physics-based and data-driven models, a physics-informed deep learning musculoskeletal model framework to learn the maps from EMGs to muscle forces and joint kinematics is proposed in this paper. The main contributions of this paper include: 1) a knowledge embedding data-driven framework is presented, which integrates the physics-based domain knowledge into the data-driven model; 2) the physics-based domain knowledge is regarded as soft constraints to penalise/regularise the loss function of deep neural networks. Physics laws relating to muscle forces and joint kinematics are applied in our case. Without loss of generality, convolutional neural network (CNN) is employed as the deep neural network to implement the proposed framework in this paper.
To validate the proposed framework for muscle forces and joint kinematics estimation, a benchmark dataset involving vast walking trials and a self-collected dataset involving wrist motion are employed.
Results indicate that the proposed framework with simpler neural network architecture outperforms selected baseline methods, including CNN, multilayer extreme learning machine (ML-ELM), support vector regression (SVR), and extreme learning machine (ELM).

The remaining of this paper is organised as follows: Methodology is detailed in Section~\ref{sec:method}, including the main framework of the proposed physics-informed deep learning method, architecture and training of CNN, and design of loss functions. Material and experimental methods are presented in Section~\ref{sec:material}. Experimental results are reported in Section~\ref{sec:results}. Finally, discussions are presented in Section~\ref{sec:discussion}, followed by conclusions in Section~\ref{sec:conclusion}.

\section{Methodology}
\label{sec:method}
In this section, we first describe the main framework of the proposed physics-informed deep learning method for musculoskeletal modelling, in the context of muscle forces and joint kinematics prediction from sEMGs. We will elaborate on the main framework, CNN architecture and training, and the designed loss functions below.

\subsection{Physics-informed Deep Learning Framework}
\begin{figure}
\centering
\includegraphics[width=1\linewidth]{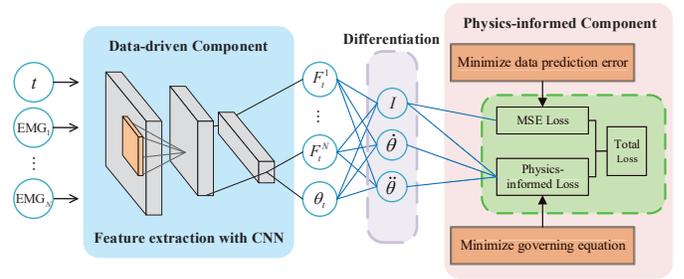}
\caption{Main framework of physics-informed deep learning.
In this study, inputs of CNN are time steps and EMG signals, while outputs of CNN are muscle forces $F^n_t$ and joint angles $\theta_t$ $\left( {n = 1, \ldots,N, t = 1, \ldots ,T} \right)$.}
\label{fig:2}
\end{figure}

Fig.~\ref{fig:2} depicts the main framework of the proposed physics-informed deep learning method for musculoskeletal modelling, in the context of muscle forces and joint kinematics prediction from sEMG.
To be specific, in the data-driven component, CNN is utilised to automatically extract the high-level features and build the relationship between EMG signals and the joint motion/muscle forces, while the physics-based component entails the underlying physical relationship between joint motion and muscle forces.
In this manner, in the data-driven component, the recorded EMG signals and the time steps are first fed into CNN.
With the features extracted by CNN, the predicted muscle forces and joint angles could be achieved. Such predictions should also satisfy the physical equation of motion, which is then taken as the soft constraint to penalise/regularise the loss function of CNN.
Finally, a modified total loss function is constructed by integrating the conventional mean square error (MSE) loss and the physics-based loss for the training purpose.

\subsection{Architecture and Training of CNN}

To demonstrate the effectiveness of the proposed physics-informed deep learning framework, a very simple architecture of CNN is employed in this paper. It only consists of one convolutional block, two fully connected blocks, and one regression block. Specifically, the convolutional block has a convolutional layer, a ReLU layer, a batch normalisation layer, and a dropout layer. It utilises the kernel size of 3, a boundary padding of 3, and a stride of 1 in the convolutional layer. There are 128 kernels in the convolutional layer and a ReLU layer is added subsequently to the convolutional layer. The batch normalisation layer is also considered to mitigate alternation made by the convolutonal layer. Similar to the convolutional block, there are one ReLU layer, one normalisation layer, and one dropout layer in each fully connected block. The number of hidden nodes is 128. Outputs of the second fully connected block are then fed into the regression block for the muscle forces and joint kinematics prediction.

In the model training phase, the batch size is set as 1, and CNN is trained by stochastic gradient descent with momentum. Additionally, the maximum iteration is 1200, and the initial learning rate is set as 0.01. The dropout rate in each dropout layer is 0.3, which means that 30\% of the hidden nodes in each layer will be randomly dropped in each iteration. Additionally, the maximum iteration is set as 1200, and the initial learning rate is set as 0.01. 

\subsection{Design of Loss Functions}
Unlike state-of-the-art methods, the loss function of the proposed framework consists of the MSE loss and physics-based loss. The MSE loss is to minimise the MSE of the ground truth and prediction, while the physics-based loss preserves the physical constraints during the movements. In this paper, the optimised CNN parameters are achieved based on the total loss below:
\begin{align}
\label{eq:total_loss}
L_{total} &=  \ L_F + L_{\theta}+ L_P \\
L_F &=  \ MSE(F)\\
L_{\theta} &=  \ MSE(\theta) \\
L_P &=  \ \Phi(F,\theta)
\end{align}
where $L_F$ denotes the prediction loss of the muscle force, while $L_{\theta}$ denotes the loss of the joint angle, respectively,
$L_P$ represents the loss function imposed by the physics law, which can penalise/regularise the loss function of a deep neural network for performance enhancement. $\Phi(F,\theta)$ denotes the function of predicted variables.
\subsubsection{MSE Loss}
MSE loss is calculated by
\begin{align}\label{eq:mse_loss}
{\rm MSE}(F) &=  \ \frac{1}{T} \displaystyle \sum_{t=1}^{T} \sum_{n=1}^{N}(F^n_t - \hat{F}^n_t)^2
 \\
{\rm MSE}(\theta) &=  \frac{1}{T} \displaystyle \sum_{t=1}^{T} (\theta_t - \hat{\theta}_t)^2
\end{align}
where $F^n_t$ represents the force of muscle $n$ at time $t$, and $\theta_t$ denotes the joint angle at time $t$, and $\hat{F}^n_t$ and $\hat{\theta}_t$ are their corresponding predicted values from the network, respectively. Additionally, $T$ denotes the total sample number, and $N$ is the number of muscles at the joint of interest.

\subsubsection{Physics-based Loss}
Physics-based governing laws, reflecting underlying relationships among the muscle force and kinematics in the human motion, are converted to constraints during the CNN training phase.
In this paper, the equation of motion is utilised to design the physics-based loss, which can be mathematically represented as
\begin{align}
\label{eq:physics_loss}
\Phi(F,\theta) =  \frac{1}{T}\sum_{t=1}^{T} (M(\theta_t) \Ddot{\theta}_t + C(\theta_t, \Dot{\theta}_t)+ G(\theta_t)  - \tau_t )^2
\end{align}
where $M(\theta_t), C(\theta_t, \Dot{\theta}_t)$, and $G(\theta_t)$ denote mass matrix, the Centrifugal and Coriolis force, and the gravity, respectively.
$\theta_t$ denotes the joint angle.
$\tau_t$ represents the joint torque, which is calculated by the summation of the product of the moment arm and muscle force:
\begin{align}
\label{eq:torque}
\tau_t = \sum_{n=1}^{N} r_{n}F^n_t
\end{align}
where $r_{n}$ is the moment arm of the muscle $n$, which is exported from OpenSim.
In this manner, along with the data-driven loss, the physics-based domain knowledge actually plays a regularisation role that enhances the robustness of the created model, and encoding such physical information into a deep neural network could also strengthen the generalisation performance of the proposed framework even when there are only a few training data.

\section{Material and Experimental Methods}
\label{sec:material}
Two datasets, including one benckmark dataset of walking trials and one self-collected dataset of wrist motions, are utilised to demonstrate the feasibility of the proposed framework.

\subsection{Benchmark Dataset}
The walking trails are retrieved from a real-world simulation~\cite{liu2008muscle}. This experiment recruited six able-bodied subjects. The mean age of these subjects was 12.9 $\pm$ 3.3 years old, and the mean weight was 51.8 $\pm$ 19.2 Kg. Subjects were informed to walking at four different walk speeds, including very slow (0.53 $\pm$ 0.04 m/s), slow (0.75 $\pm$ 0.10 m/s), free (1.15 $\pm$ 0.08 m/s), and fast speeds (1.56 $\pm$ 0.21 m/s).

The data, including GRFs and markers' data, were imported to OpenSim to scale the generic musculoskeletal model for each subject~\cite{seth2018opensim}.
The joint kinematics and joint torque were computed through the inverse kinematic (IK) an inverse dynamic (ID) tools, respectively.
The muscle forces were computed using the computed muscle control (CMC) tool to ensure the muscle excitations followed the measured EMGs~\cite{thelen2003generating}.
Each gait cycle was normalised into 100 frames.
We opt to estimate the joint angle and muscle forces at the knee joint during different walking speeds from the EMGs.
The $biceps~femoris~short~head$ (BFS) and the $retus~femoris$ (RF) are chosen as they are the main flexor and extensor of knee joint~\cite{ma2016patient}.

Each walking trial is formed to a 100-by-7 matrix consisting of the time step, gait cycle, the enveloped EMG signals, and BFS and RF muscle forces. All walking trials are concatenated for each subject to form a single long matrix for the proposed framework.

\subsection{Self-Collected Dataset}
Approved by the MaPS and Engineering Joint Faculty Research Ethics Committee of the University of Leeds (MEEC 18-002), six subjects were recruited to participate in this experiment. All subjects gave signed consent.
In the experiment, subjects were informed to maintain a fully straight torso with the $90^{\circ}$ abducted shoulder and the $90^{\circ}$ flexed elbow joint. The continuous wrist flexion/extension motion was recorded using the VICON motion capture system.  The joint motions were computed through the upper limb model using 16 reflective markers (sampled at~\SI{250}{Hz}).
Meanwhile, EMG signals were recorded by Avanti Sensors (sampled at~\SI{2000}{Hz}) from the main wrist muscles ($n = 1,2,\ldots,5$), including the $flexor~carpi~radialis$ (FCR), the $flexor~carpi~ulnaris$ (FCU), the $extensor~carpi~radialis~longus$ (ECRL), the $extensor~carpi~radialis~brevis$ (ECRB), and the $extensor~carpi~ulnaris$ (ECU).
The electrodes were allocated by palpation and evaluated by performing contraction while looking at the signal before the experiment. Moreover, the EMG signals and motion data were synchronised and resampled at \SI{1000}{Hz}. Five repetitive trials were performed for each subject, and a three-minute break was given between trials to prevent muscle fatigue.

The measured EMG signals were band-pass filtered (\SI{20}{Hz} and~\SI{450}{Hz}), fully rectified, and low-pass filtered (\SI{6}{Hz}). Then, they were normalised concerning the maximum voluntary contraction recorded before the experiment, resulting in the enveloped EMG signal. The markers' data are used to compute the wrist kinematics via the IK tool. Then the joint torque and wrist muscle forces are obtained from the ID and CMC tools ensured the computed motion was consistent with the measured joint motion.

Each wrist motion trial, consisting of time steps, filtered EMG signals, wrist muscle forces, and wrist joint angles, is formed into a $t$ by 12 matrix.

\subsection{Baseline Methods and Parameters Setting}
To verify the effectiveness of the proposed physics-informed deep learning framework, several state-of-the-art methods, including CNN, multilayer extreme learning machine (ML-ELM)~\cite{zhang2020non}, support vector regression (SVR), and ELM~\cite{huang2011extreme}, are considered as the baseline methods for the comparison. Specifically, CNN has three convolutional blocks, three fully connected blocks, and one regression block is considered in the experiments. Stochastic gradient descent with momentum optimiser is employed for CNN training; the batch size is set as 1, the maximum iteration is set as 1200, and the initial learning rate is 0.01. ML-ELM has five hidden layers, and the number of hidden nodes in each hidden layer is determined by the grid search method. ELM is a single hidden layer feedforward neural network, and the sigmoid function is utilised as the activation function in this paper.

\subsection{Evaluation Criteria}
To quantify the estimation performance of the proposed framework, root mean square error (RMSE) is first used as the metric. In specific, RMSE indicates the discrepancies in the amplitude and between the estimated variables and ground truth, which can be calculated by
\begin{align}
\label{eq:nrmse}
{\rm RMSE} =  \sqrt{\frac{1}{T}\sum_{t=1}^{T}(y_t- \hat y_t)^2}
\end{align}
where $y_t$ and $\hat y_t$ indicate the ground truth and the corresponding predicted value, respectively.

Pearson's correlation coefficient ($CC$) is also employed as another metric, which could be calculated by
\begin{align}
\label{eq:r}
CC= \frac{\displaystyle \sum_{t=1}^{T}(y_t - \overline{y_t})(\hat y_t - \overline{\hat y_t})}{\sqrt{\displaystyle \sum_{t=1}^{T}(y_t - \overline{y_t})^2}\sqrt{\displaystyle \sum_{t=1}^{T}(\hat y_t - \overline{\hat y_t})^2}}
\end{align}
where $\overline{y_t}$ and $\overline{\hat y_t}$ are the mean of the ground truth and predicted value, respectively.

\section{Results}
\label{sec:results}
In this section, we verify the performance of the proposed framework on knee joint and wrist joint scenarios via comparing with selected baseline methods. Specifically, the training process of the proposed framework is first illustrated. Over comparisons are then performed to demonstrate the predicted results of the proposed framework and baseline methods, including representative results of predicted muscles and joint angles, and detailed and average predicted results of six able-bodied subjects. Finally, the intrasession scenario is also considered to evaluate the robustness and generalisation performance of the proposed framework. The training of the proposed framework and baseline methods is carried out on a workstation with GeForce RTX 2080 Ti graphic cards and 128G RAM.

\subsection{Training Process of the Proposed Framework}
\begin{figure}
\centering
\includegraphics[width=1\linewidth]{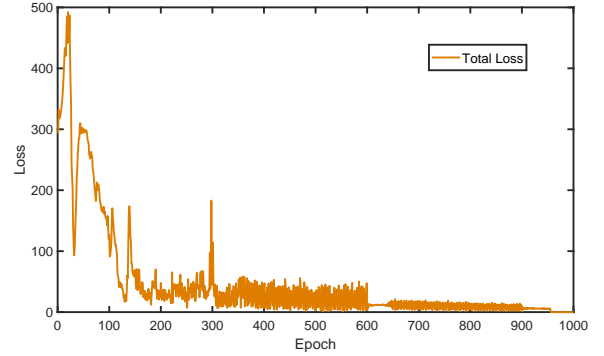}
\caption{Illustration of the total loss of wrist joint scenario.}
\label{fig:4}
\end{figure}

To demonstrate the convergence of the proposed framework, we illustrate the convergence process of the total loss of wrist joint case during the training phase in Fig.~\ref{fig:4} as the exemplar. For the wrist angle and muscle forces prediction of the wrist joint scenario, the total loss is very low after 200 iterations with small local oscillations as the iteration progresses and actually converges after 600 iterations. We conjecture that the main reason is that we set the batch size as 1 during training CNN. Such batch size could help CNN learn the data distribution better, but it also causes local oscillations. We also would like to point out that for the knee joint case, the same convergence process has also been achieved.

\begin{figure*}
\centering
\begin{minipage}{0.32\linewidth}
\centering
\includegraphics[width=1.05\linewidth]{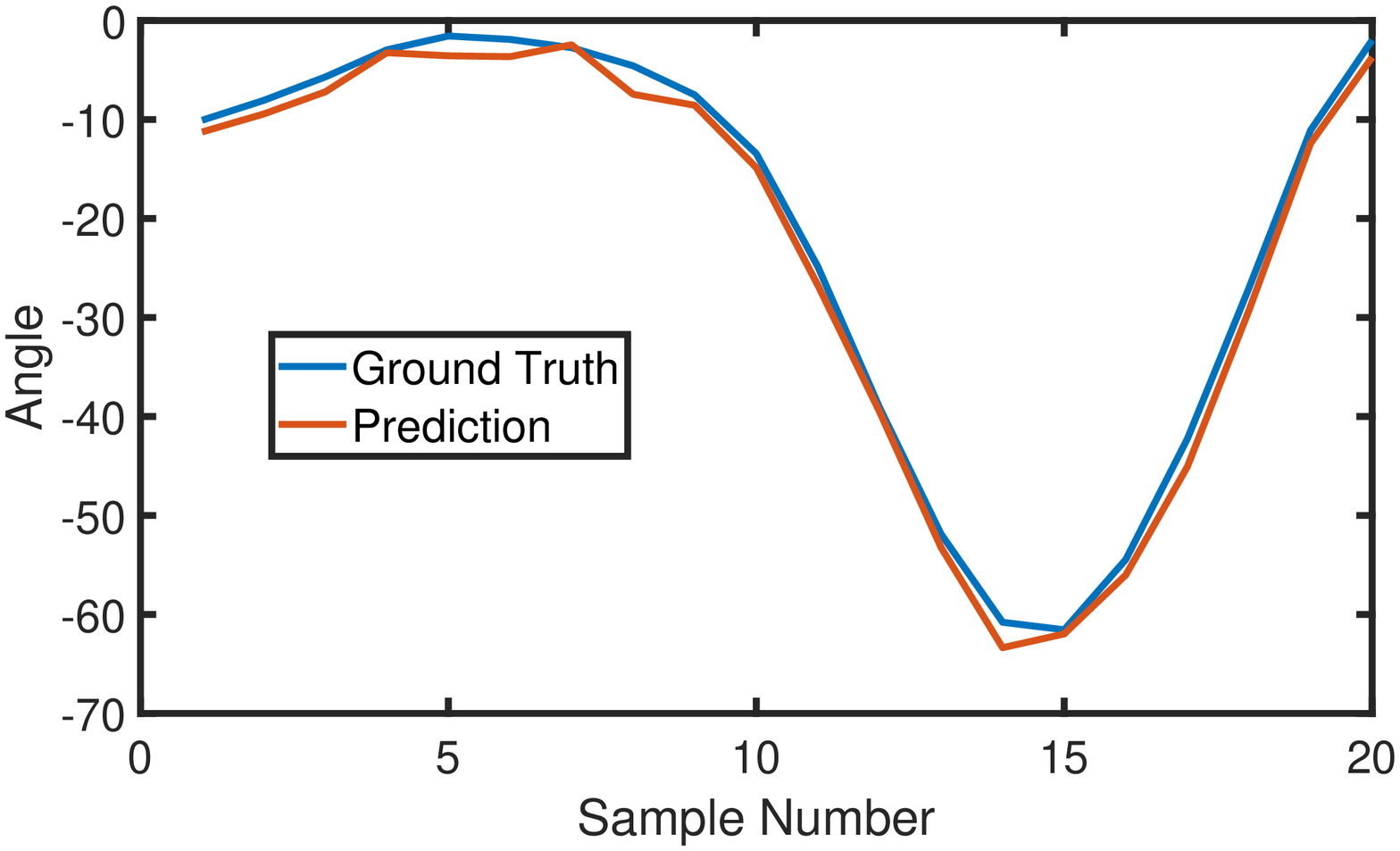}
\centerline{{\fontsize{7.5pt}{9.8pt}\selectfont (a) Predicted results of knee angle}}
\end{minipage}
\begin{minipage}{0.32\linewidth}
\centering
\includegraphics[width=1.05\linewidth]{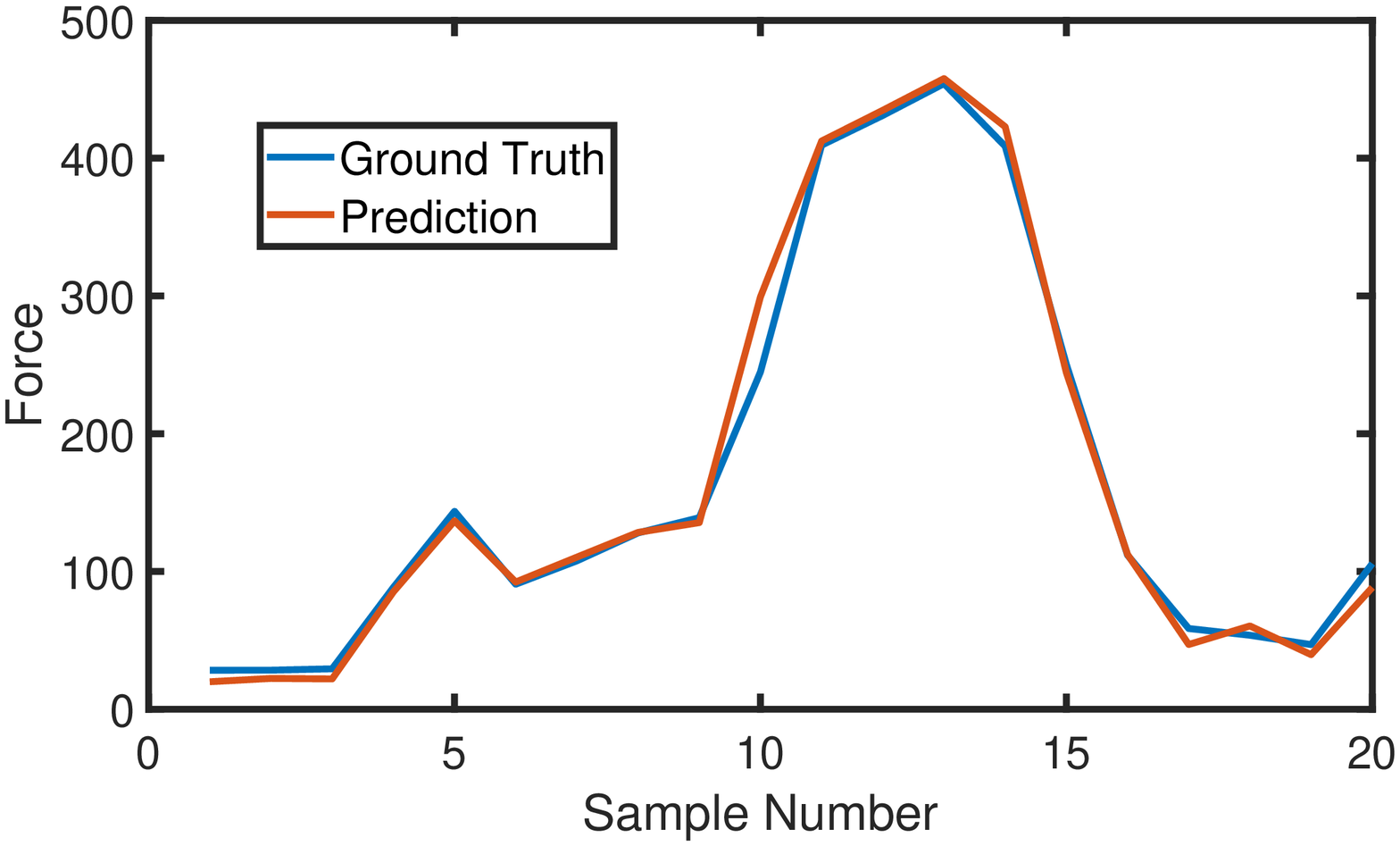}
\centerline{{\fontsize{7.5pt}{9.8pt}\selectfont (b) Predicted results of BFS}}
\end{minipage}
\begin{minipage}{0.32\linewidth}
\centering
\includegraphics[width=1.05\linewidth]{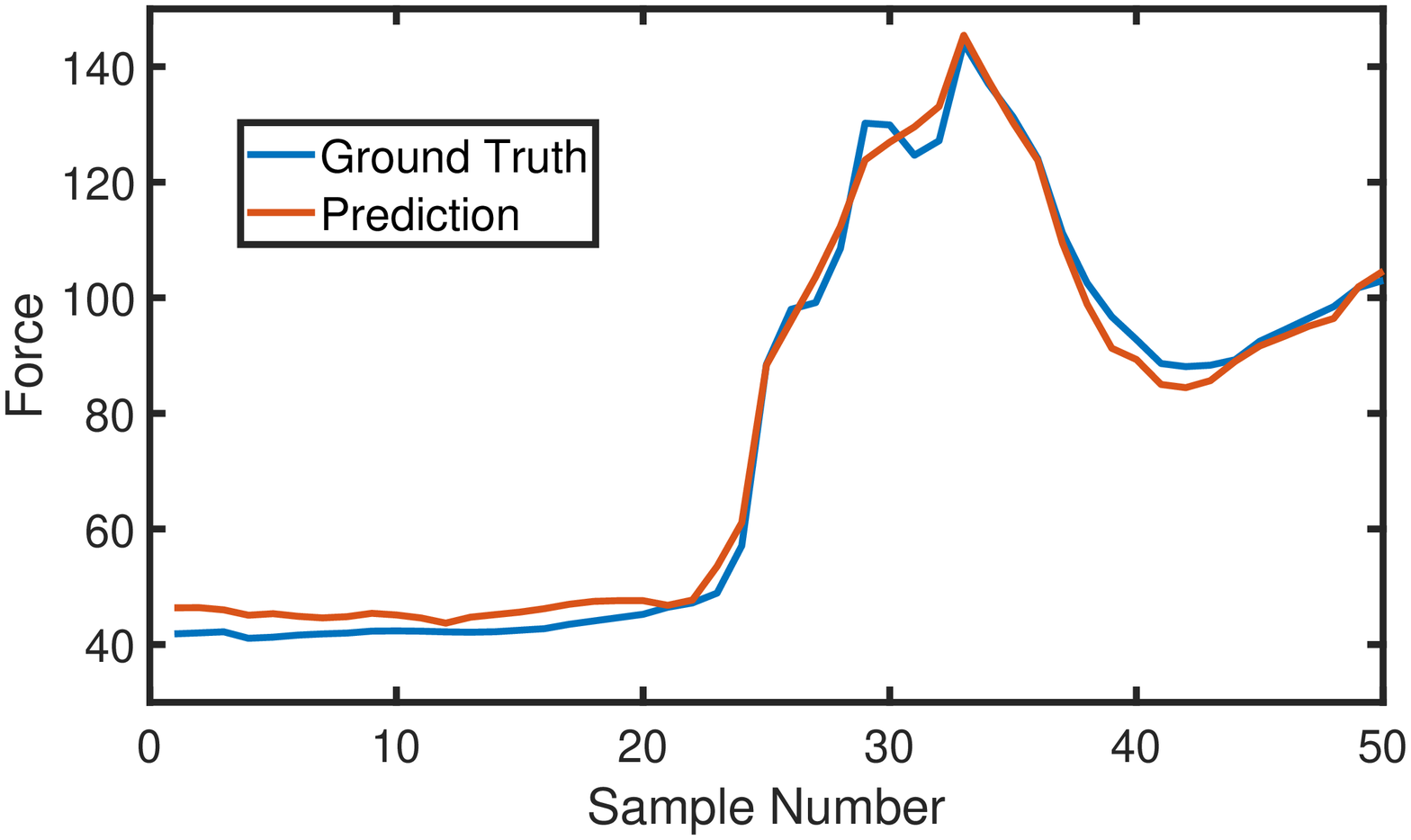}
\centerline{{\fontsize{7.5pt}{9.8pt}\selectfont (c) Predicted results of RF}}
\end{minipage}\\
\caption{Representative results of the knee joint through proposed physics-informed data-driven model. The predicted outputs of the knee joint case include the knee angle, BFS muscle force, and RF muscle force.}
\label{fig:6}
\end{figure*}

\subsection{Overall Comparisons}
\begin{figure*}
\centering
\begin{minipage}{0.32\linewidth}
\centering
\includegraphics[width=1.05\linewidth]{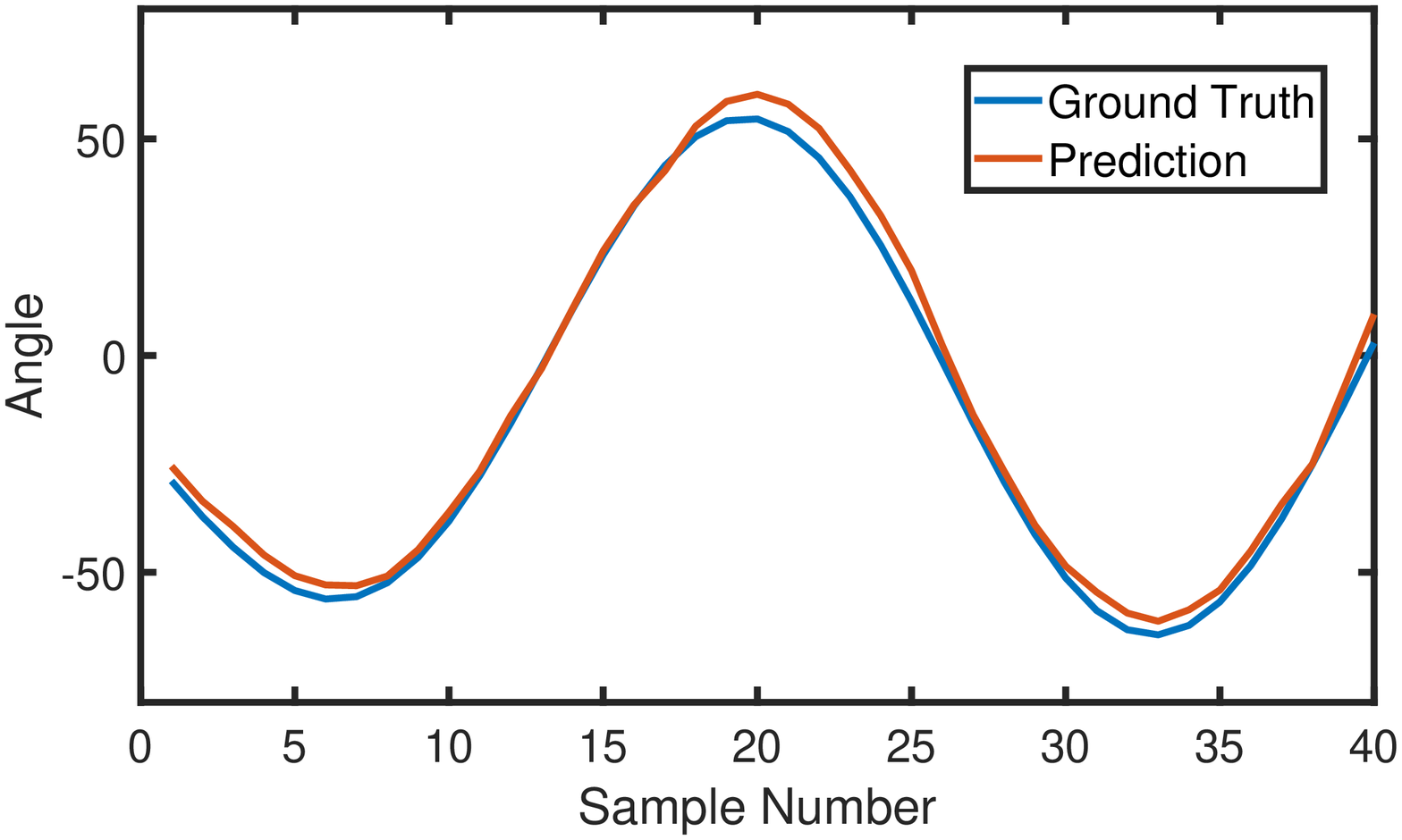}
\centerline{{\fontsize{7.5pt}{9.8pt}\selectfont (a) Predicted results of wrist angle}}
\end{minipage}
\begin{minipage}{0.32\linewidth}
\centering
\includegraphics[width=1.05\linewidth]{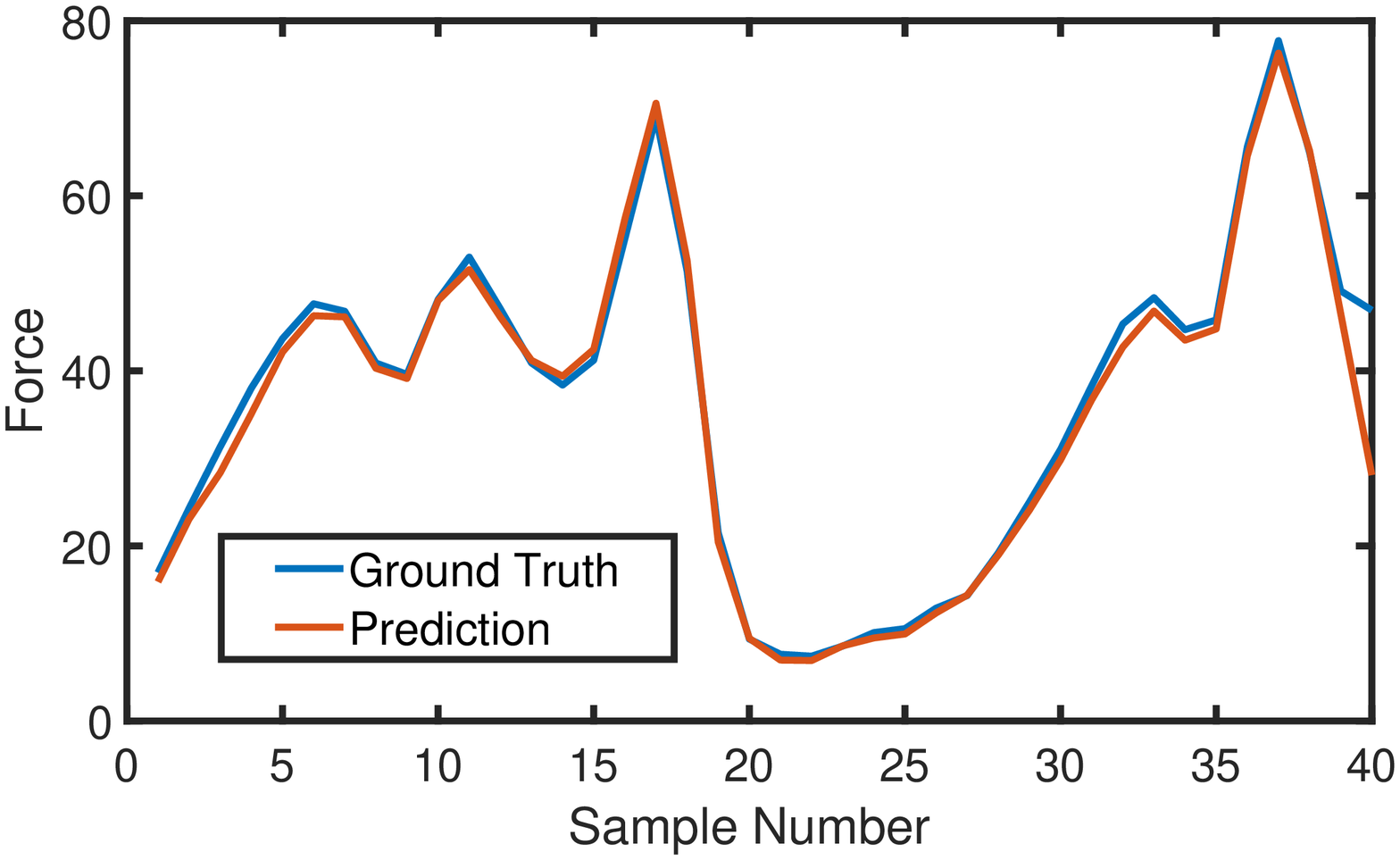}
\centerline{{\fontsize{7.5pt}{9.8pt}\selectfont (b) Predicted results of FCR}}
\end{minipage}
\begin{minipage}{0.32\linewidth}
\centering
\includegraphics[width=1.05\linewidth]{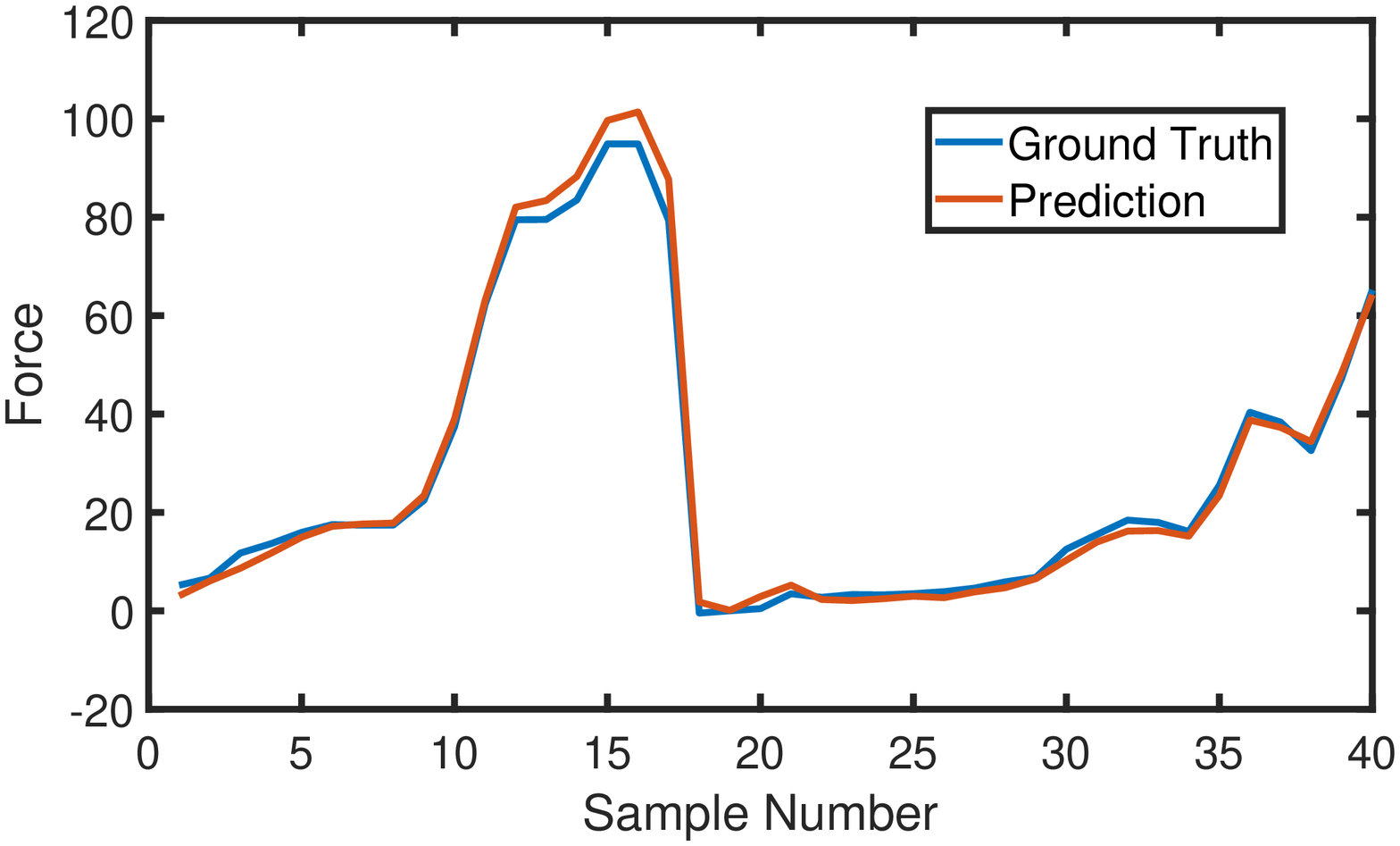}
\centerline{{\fontsize{7.5pt}{9.8pt}\selectfont (c) Predicted results of FCU}}
\end{minipage}\\
\begin{minipage}{0.32\linewidth}
\centering
\includegraphics[width=1.05\linewidth]{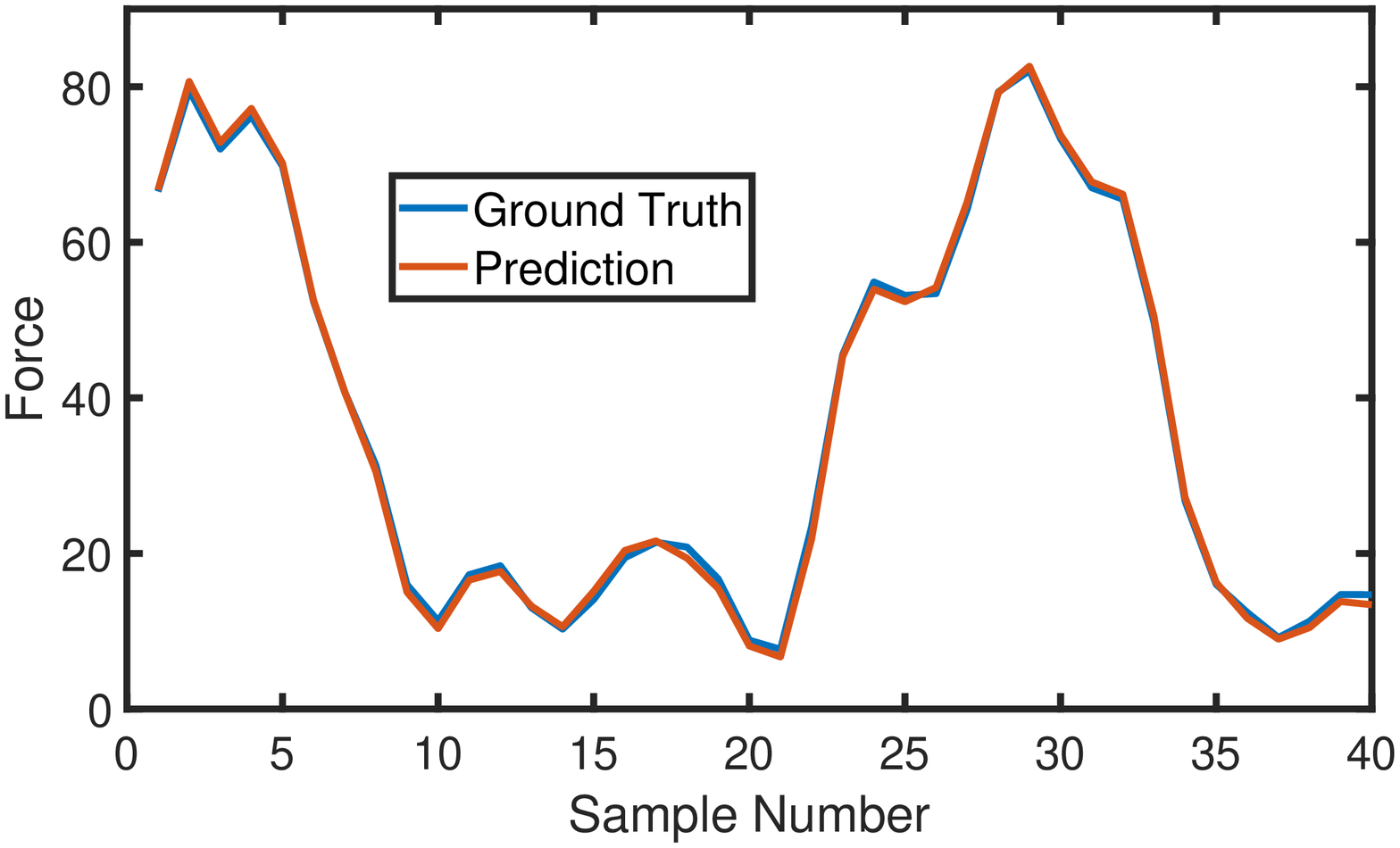}
\centerline{{\fontsize{7.5pt}{9.8pt}\selectfont (d) Predicted results of ECRL}}
\end{minipage}
\begin{minipage}{0.32\linewidth}
\centering
\includegraphics[width=1.05\linewidth]{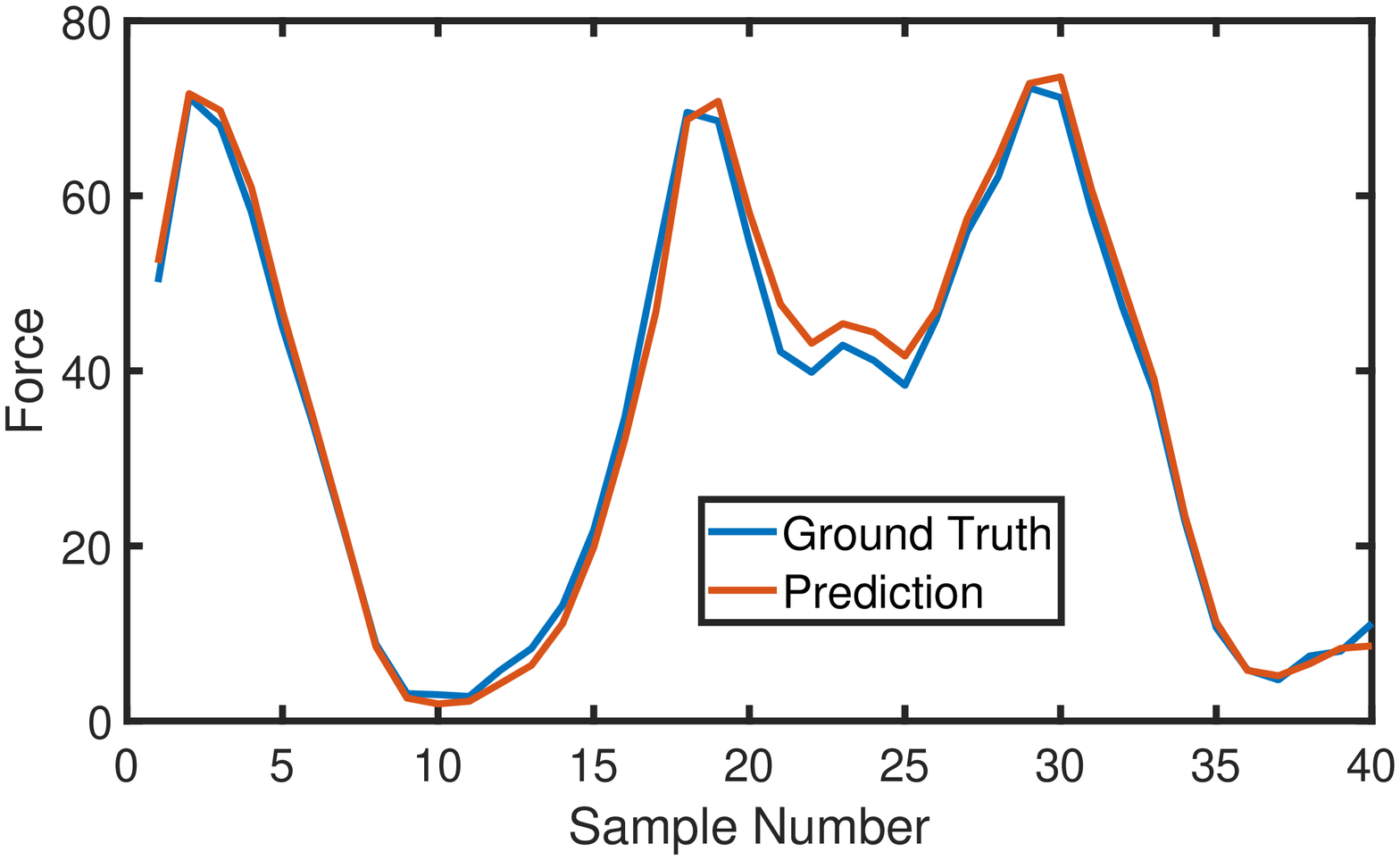}
\centerline{{\fontsize{7.5pt}{9.8pt}\selectfont (e) Predicted results of ECRB}}
\end{minipage}
\begin{minipage}{0.32\linewidth}
\centering
\includegraphics[width=1.05\linewidth]{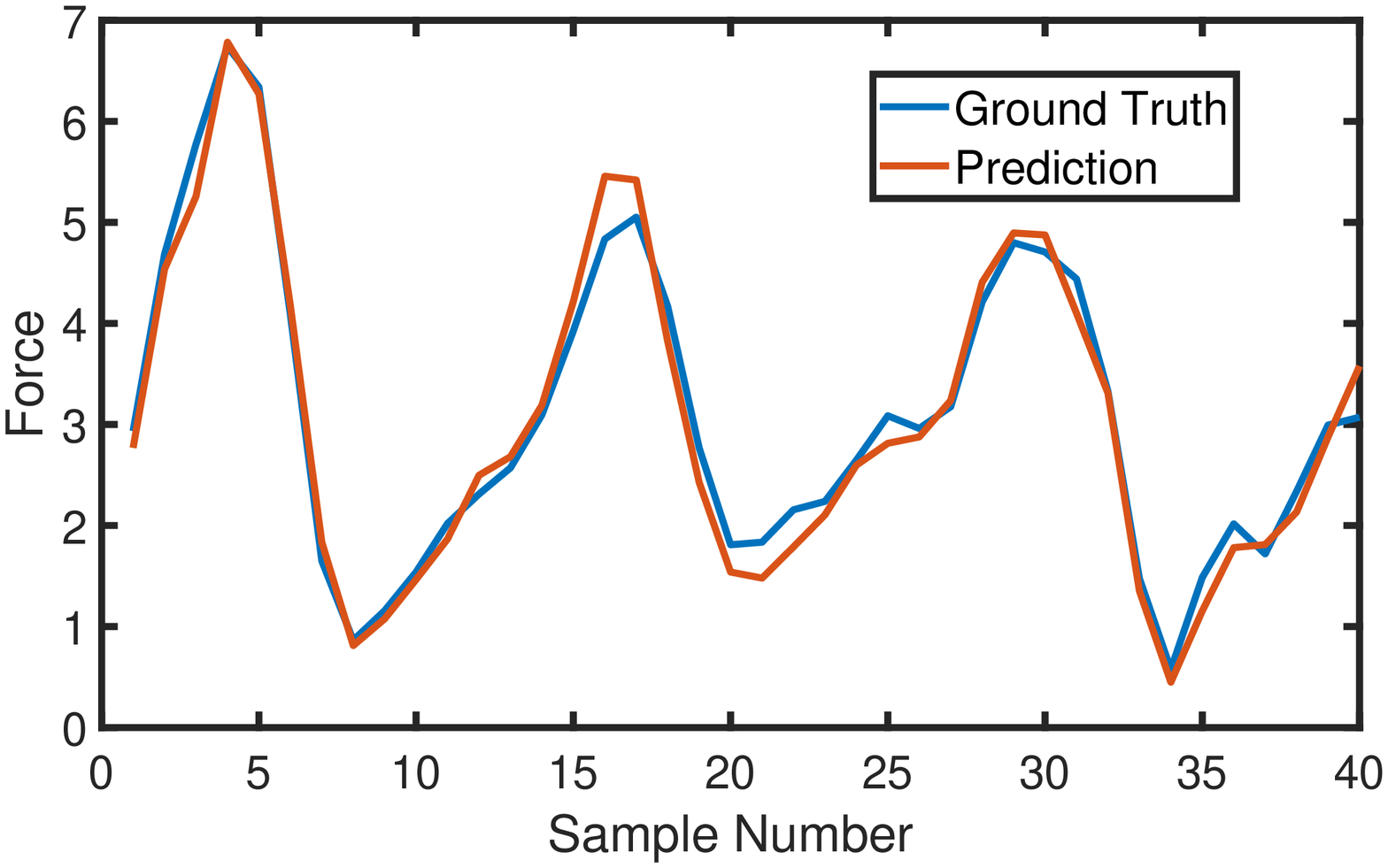}
\centerline{{\fontsize{7.5pt}{9.8pt}\selectfont (f) Predicted results of ECU}}
\end{minipage}\\
\caption{Representative results of the wrist joint through the knowledge embedding data-driven model. The predicted outputs of the wrist joint case include the wrist angle, FCR muscle force, FCU muscle force, ECRL muscle force, ECRB muscle force, and ECU muscle force.}
\label{fig:7}
\end{figure*}

The overall comparisons between the proposed framework and baseline methods are first performed. Fig.~\ref{fig:6} and Fig.~\ref{fig:7} depict the representative results of the proposed framework for knee joint and wrist joint, including knee flexion angle, muscle force of RF, muscle force of BFS, wrist flexionangle, muscle force of FCR, muscle force of FCU, muscle force of ECRL, muscle force of ECRB, and muscle force of ECU, respectively. As we can see from Fig.~\ref{fig:6} and Fig.~\ref{fig:7}, the predicted values of muscle forces and joint angles could fit the ground truths well, indicating the great dynamic tracking capability of the proposed framework.

To quantitatively evaluate the performance of the proposed framework, detailed comparisons of all the subjects between the proposed framework and baseline methods are presented in Table~\ref{tab:1} and Table~\ref{tab:2}. The proposed framework could achieve smaller RMSEs and higher Pearson correlation coefficients in most cases, which further verifies the robustness of the proposed framework. To be specific, deep learning-based methods, including the proposed framework, CNN and ML-ELM, achieve better predicted performance than machine learning-based methods, including SVR and ELM. Because these deep learning-based methods could automatically extract high-level features from the collected data. Among deep learning-based methods, the proposed framework achieves the best predicted performance, because the embedded physics law could penalise/regularise the CNN in the proposed framework, its performance is not only dependent on the conventional MSE loss, but also can be enhanced by the physics-based loss.

\begin{table*}[htbp]
    \caption{RMSE and $CC$ of the proposed framework and baseline methods of knee joint case}
    \label{tab:1}
    \centering
    \resizebox{0.7\linewidth}{!}
    {
    \begin{tabular}{c|c|ccc|c|c|ccc}
    \toprule
    \bfseries Subject & \bfseries Methods  & \bfseries RF($N$)& \bfseries BFS($N$)& \bfseries Knee Angle($^\circ$) & \bfseries Subject & \bfseries Methods & \bfseries RF($N$)& \bfseries BFS($N$)& \bfseries Knee Angle($^\circ$)\\
    \midrule
    \multirow{5}{*}{\bfseries S1} & Ours & 18.19/0.92 & 14.16/0.99 & 5.69/0.99 & \multirow{5}{*}{\bfseries S4} & Ours & 21.51/0.95 & 15.95/0.99 & 7.07/0.99 \\
    \cmidrule{2-5} \cmidrule{7-10}
    &   CNN& 20.04/0.89 & 9.75/0.99 & 2.17/0.97 && CNN & 31.66/0.94 & 16.34/0.99 & 7.59/0.99 \\
    \cmidrule{2-5} \cmidrule{7-10}
    &   ML-ELM & 25.31/0.86 & 17.96/0.92 & 6.01/0.97 & & ML-ELM & 29.74/0.94 & 19.61/0.99 & 12.39/0.97 \\
    \cmidrule{2-5} \cmidrule{7-10}
    &   ELM & 35.93/0.82 & 23.85/0.90 & 16.82/0.94 &  & ELM & 35.80/0.91 & 26.58/0.93 & 15.97/0.93 \\
    \cmidrule{2-5} \cmidrule{7-10}
    &   SVR & 33.20/0.81 & 27.99/0.90 & 15.33/0.93 &  & SVR & 33.29/0.92 & 23.23/0.93 & 14.74/0.97 \\
    \midrule
    \multirow{5}{*}{\bfseries S2} & Ours & 17.15/0.94 & 19.83/0.93 & 4.30/0.99 & \multirow{5}{*}{\bfseries S5} & Ours & 13.24/0.97 & 15.52/0.94 & 4.71/0.99 \\
    \cmidrule{2-5} \cmidrule{7-10}
    &   CNN& 23.96/0.92 & 20.95/0.94 & 3.55/0.98 &  & CNN & 15.31/0.95 & 15.39/0.94 & 6.97/0.99 \\
    \cmidrule{2-5} \cmidrule{7-10}
    &   ML-ELM & 21.37/0.91 & 29.62/0.91 & 9.28/0.94 && ML-ELM  & 19.26/0.95 & 20.28/0.90 & 6.05/0.99 \\
    \cmidrule{2-5} \cmidrule{7-10}
    &   ELM & 29.21/0.88 & 36.25/0.83 & 17.66/0.91 &  & ELM & 30.98/0.90 & 23.92/0.91 & 10.22/0.96 \\
    \cmidrule{2-5} \cmidrule{7-10}
    &   SVR & 31.03/0.89 & 31.53/0.85 & 14.31/0.90 &  & SVR & 25.73/0.89 & 25.37/0.92 & 11.36/0.97 \\
    \midrule
    \multirow{5}{*}{\bfseries S3} & Ours & 15.51/0.94 & 13.27/0.92 & 5.13/0.98 & \multirow{5}{*}{\bfseries S6} & Ours & 18.71/0.95 & 15.47/0.95 & 5.63/0.98 \\
    \cmidrule{2-5} \cmidrule{7-10}
    &   CNN& 13.20/0.96 & 17.56/0.92  & 4.25/0.99 & & CNN & 19.73/0.96 & 12.55/0.94 & 7.37/0.98 \\
    \cmidrule{2-5} \cmidrule{7-10}
    &   ML-ELM & 16.77/0.93 & 21.38/0.93& 7.99/0.97 && ML-ELM  & 25.36/0.93 & 21.34/0.92& 8.96/0.97 \\
    \cmidrule{2-5} \cmidrule{7-10}
    &   ELM & 26.35/0.90 & 26.78/0.90 & 19.66/0.94&  & ELM & 29.75/0.93 & 22.53/0.92 & 12.31/0.94 \\
    \cmidrule{2-5} \cmidrule{7-10}
    &   SVR & 24.60/0.89 & 22.97/0.90 & 9.39/0.97 &  & SVR & 27.38/0.88 & 28.70/0.89 & 15.29/0.95 \\
    \bottomrule
    \end{tabular}}
\end{table*}

\begin{table*}[htbp]
    \caption{RMSE and $CC$ of the proposed framework and baseline methods of wrist joint case}
    \label{tab:2}
    \centering
    \resizebox{\linewidth}{!}
    {
    \begin{tabular}{c|c|cccccc|c|c|cccccc}
    \toprule
    \bfseries Subject & \bfseries Methods  & \bfseries FCR($N$)& \bfseries FCU($N$)& \bfseries ECRL($N$) & \bfseries ECRB($N$) & \bfseries ECU($N$) & \bfseries Wrist Angle($^\circ$) & \bfseries Subject & \bfseries Methods  & \bfseries FCR($N$)& \bfseries FCU($N$)& \bfseries ECRL($N$) & \bfseries ECRB($N$) & \bfseries ECU($N$)& \bfseries Wrist Angle($^\circ$)\\
    \midrule
    \multirow{5}{*}{\bfseries S1} & Ours & 3.25/0.99 & 2.51/0.98 & 0.79/0.99 & 2.21/0.99 & 0.24/0.98 & 3.76/0.99& \multirow{5}{*}{\bfseries S4} & Ours & 3.91/0.98 & 2.79/0.99 & 0.57/0.99 & 3.26/0.97 & 0.33/0.99 & 4.31/0.97 \\
    \cmidrule{2-8} \cmidrule{10-16}
    & CNN & 2.81/0.99 & 2.32/0.99 & 0.90/0.99 & 1.66/0.99 & 0.16/0.99 & 2.30/0.99 & & CNN & 2.89/0.99 & 3.03/0.99 & 0.81/0.99 & 3.36/0.98 & 0.20/0.99 & 4.25/0.97\\
    \cmidrule{2-8} \cmidrule{10-16}
    & ML-ELM & 5.78/0.97 & 3.10/0.98 & 3.51/0.97 & 3.03/0.96 & 0.69/0.99 & 6.98/0.98 & & ML-ELM & 6.33/0.98 & 4.93/0.97 & 1.38/0.99 & 4.90/0.97 & 0.88/0.99 & 5.89/0.97\\
    \cmidrule{2-8} \cmidrule{10-16}
    & ELM & 10.55/0.93 & 6.21/0.95 & 4.77/0.96 & 3.87/0.96 & 1.73/0.98 & 12.33/0.93 & & ELM & 8.27/0.96 & 7.38/0.95 & 3.55/0.97 & 5.73/0.95 & 1.03/0.99 & 10.55/0.94\\
    \cmidrule{2-8} \cmidrule{10-16}
    & SVR & 6.34/0.97 & 7.92/0.95 & 5.32/0.93 & 5.10/0.96 & 0.99/0.97 & 8.57/0.94 & & SVR & 9.36/0.95& 7.62/0.95 & 4.30/0.97 & 5.46/0.96 & 1.52/0.99 & 9.33/0.93\\
    \midrule
     \multirow{5}{*}{\bfseries S2} & Ours & 4.21/0.99 & 2.63/0.99 & 0.71/0.99 & 3.25/0.98 & 0.58/0.99 & 2.77/0.99 & \multirow{5}{*}{\bfseries S5} & Ours & 2.53/0.98 &3.52/0.99 &1.21/0.98 &2.91/0.99 &0.65/0.99 &3.45/0.99 \\
    \cmidrule{2-8} \cmidrule{10-16}
    & CNN & 4.29/0.99 & 3.96/0.99 & 0.99/0.99 & 2.98/0.99 & 0.37/0.99 & 2.59/0.99 & & CNN &2.31/0.98 &3.49/0.98 & 1.79/0.98 & 2.03/0.99 & 0.71/0.99 & 2.99/0.99 \\
    \cmidrule{2-8} \cmidrule{10-16}
    & ML-ELM & 7.30/0.98 & 3.27/0.98 & 2.38/0.97 & 3.57/0.99 & 0.95/0.99 & 4.30/0.96 & & ML-ELM & 6.38/0.94 & 4.33/0.97 & 3.59/0.98 & 3.87/0.98 & 1.01/0.99 & 7.56/0.96\\
    \cmidrule{2-8} \cmidrule{10-16}
    & ELM & 11.25/0.94 & 7.99/0.98 & 2.95/0.97 & 7.98/0.98 & 2.09/0.99 & 7.62/0.96 & & ELM & 9.22/0.94 & 6.27/0.93 & 8.26/0.97 & 5.66/0.98 & 2.89/0.99 & 10.30/0.92\\
    \cmidrule{2-8} \cmidrule{10-16}
    & SVR & 11.03/0.94 & 9.28/0.98 & 4.17/0.94 & 6.22/0.98 & 1.28/0.99 & 6.53/0.97 & & SVR & 8.89/0.93 & 9.35/0.92 & 6.30/0.96 & 7.27/0.97 & 2.50/0.99 & 11.28/0.93\\
    \midrule
    \multirow{5}{*}{\bfseries S3} & Ours & 5.18/0.98 & 3.77/0.98 & 0.97/0.99 & 4.96/0.97 & 0.41/0.99 & 4.30/0.97 & \multirow{5}{*}{\bfseries S6} & Ours & 6.20/0.98 & 4.17/0.98 & 0.91/0.99 & 3.89/0.99 & 0.33/0.99 & 5.81/0.97 \\
    \cmidrule{2-8} \cmidrule{10-16}
    & CNN & 3.99/0.99 & 4.59/0.98 & 1.23/0.99 & 6.83/0.98 & 0.47/0.99 & 3.94/0.98 & & CNN & 4.27/0.98 & 5.35/0.96 & 0.99/0.99 & 5.36/0.97 & 0.56/0.99 & 6.21/0.98\\
    \cmidrule{2-8} \cmidrule{10-16}
    & ML-ELM & 7.89/0.96 & 4.51/0.97 & 4.37/0.98 & 7.31/0.96 & 0.92/0.99 & 6.88/0.99 & & ML-ELM & 6.34/0.97 & 7.95/0.96 & 2.57/0.99 & 7.90/0.95 & 0.79/0.99 & 8.30/0.96\\
    \cmidrule{2-8} \cmidrule{10-16}
    & ELM & 12.33/0.93 & 8.26/0.95 & 6.25/0.95 & 7.15/0.96  & 2.02/0.99 & 9.37/0.94 & & ELM & 10.28/0.94 & 9.21/0.96 & 4.43/0.98 & 5.21/0.95 & 2.37/0.98 & 10.22/0.96\\
    \cmidrule{2-8} \cmidrule{10-16}
    & SVR & 11.76/0.94 & 7.93/0.95 &  7.01/0.95 & 8.89/0.94  & 1.78/0.98 & 7.55/0.93 & & SVR & 10.04/0.94 & 8.89/0.94 & 4.21/0.98 & 5.12/0.96 & 2.17/0.99 & 8.95/0.97\\
    \bottomrule
    \end{tabular}}
\end{table*}

\begin{figure}
\centering
\begin{minipage}{1\linewidth}
\centering
    \includegraphics[width=1\linewidth]{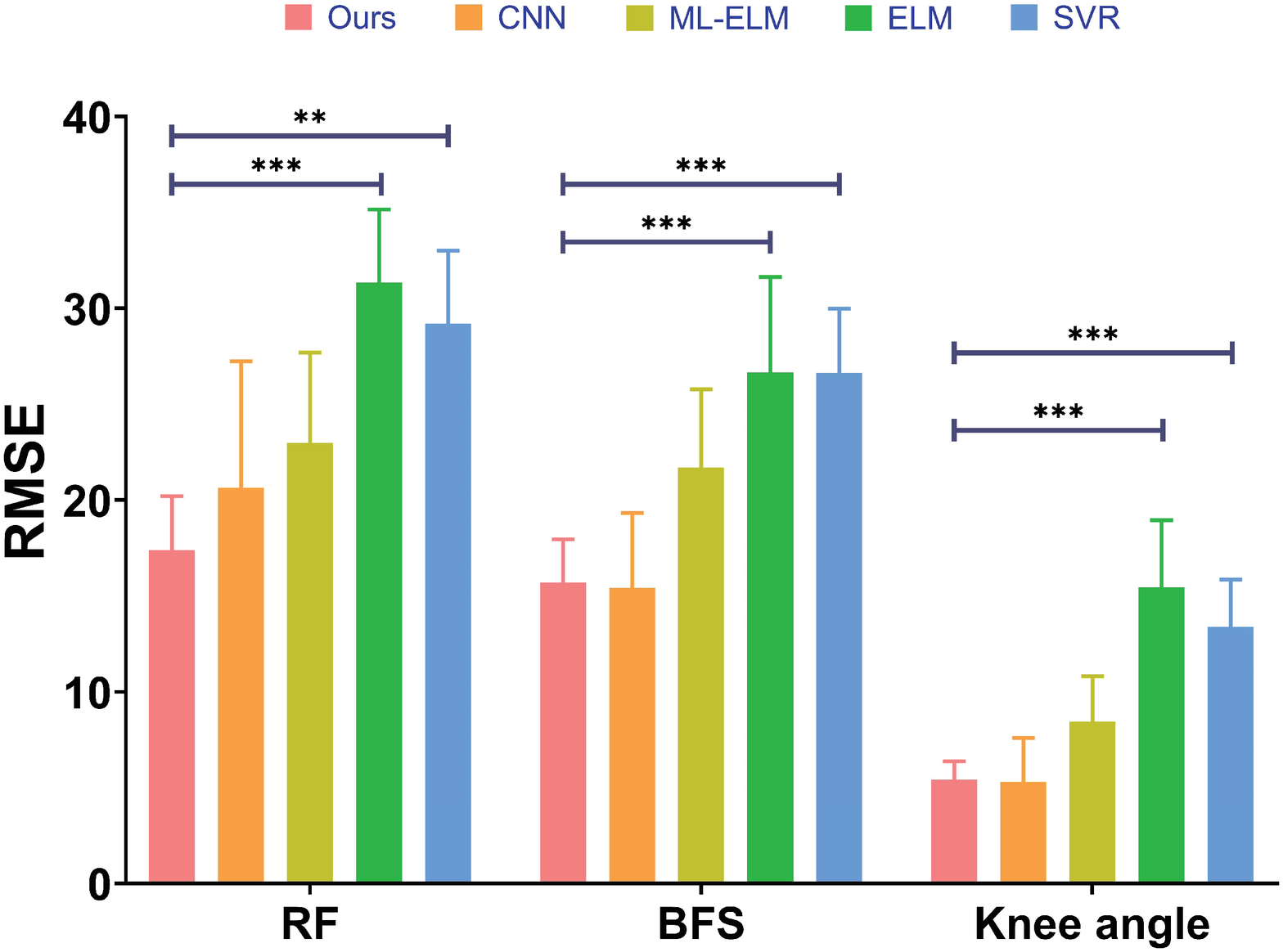}
{\fontsize{7.5pt}{9.8pt}\selectfont (a) Average RMSEs of knee joint case}
\end{minipage}\\
\begin{minipage}{1\linewidth}
\centering
\includegraphics[width=1\linewidth]{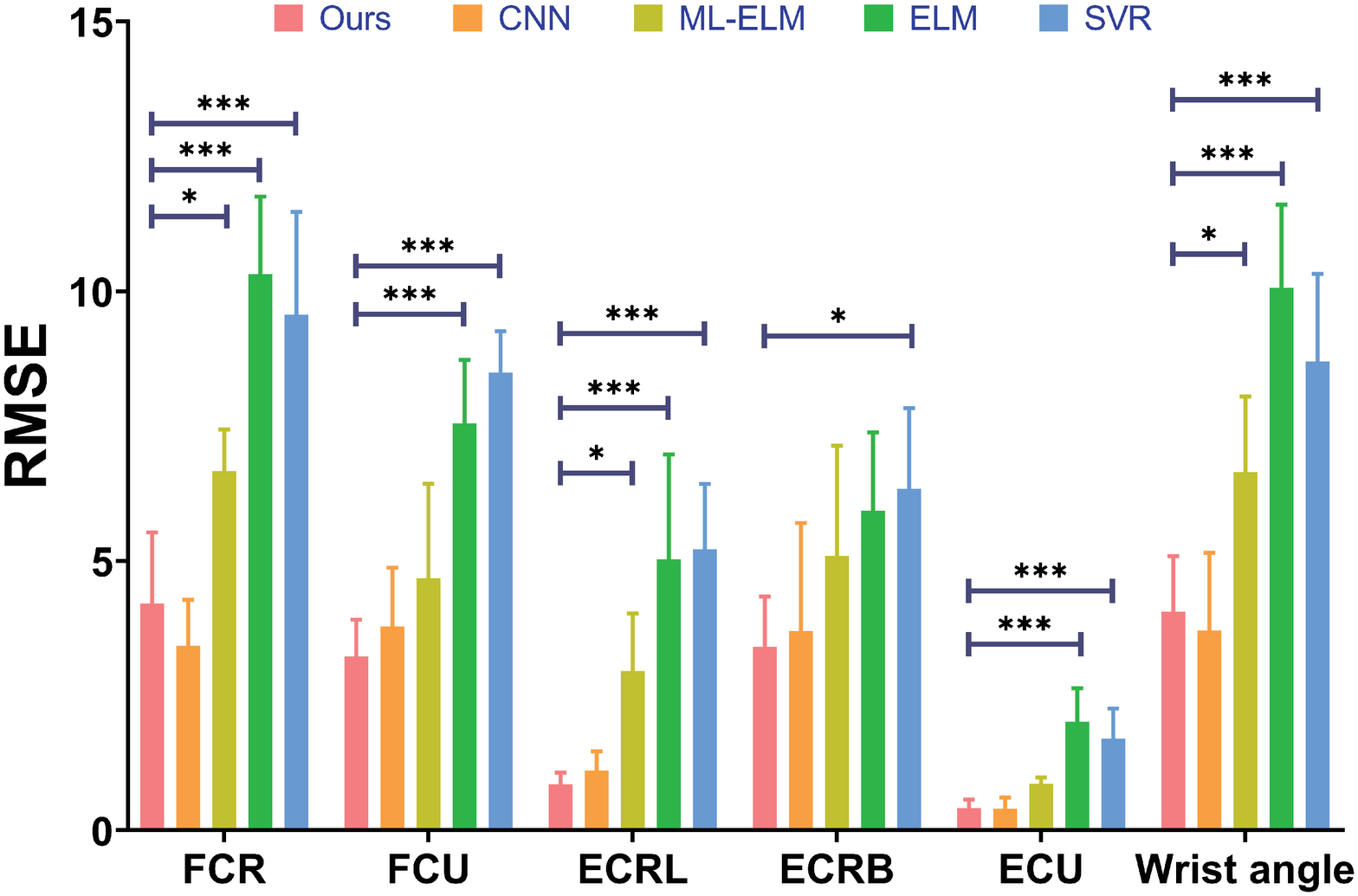}
{\fontsize{7.5pt}{9.8pt}\selectfont (b) Average RMSEs of wrist joint case}
\end{minipage}\\
\caption{Average RMSEs across all the subjects in (a) knee joint and (b) wrist joint scenarios, respectively. The proposed framework achieves comparable RMSEs with simpler CNN architecture compared with pure CNN. Compared with ML-ELM, ELM and SVR, the proposed framework achieves better and more stable prediction performance. The significance level is set as 0.05 ($^{***}p < 0.001, ^{**}p < 0.01, and   ^{*}p < 0.05)$.}
\label{fig:8}
\end{figure}

Aside from the overall comparison of the five approaches, we further carried out a pairwise analysis between the proposed method and each comparison method on the two datasets.
One-way analysis of variance (ANOVA) is conducted for statistical analysis of the proposed framework and baseline methods. RMSE is the response variable. A \textit{post-hoc} analysis using Tukey's Honest Significant Difference test is applied. The significance level is set at $p < 0.05$.
Fig.~\ref{fig:8} illustrates the average RMSEs of the knee and wrist joints of the proposed framework and baseline methods across all the subjects. As observed from Fig.~\ref{fig:8}, the proposed framework achieves satisfactory performance with lower standard deviations, and its predicted results are with smaller fluctuations. Moreover, with simple neural network architecture, the proposed framework could achieve comparable performance compared with pure CNN, which has a more complex neural network architecture, by embedding the underlying physical interactions between predicted variables into the data-driven model.
This motivates us to employ more constraints to deeply integrate the knowledge of the musculoskeletal model into the deep neural network to enhance the performance in future work.

\subsection{Evaluation of Intrasession Scenario}
\begin{figure}
\centering
\includegraphics[width=1\linewidth]{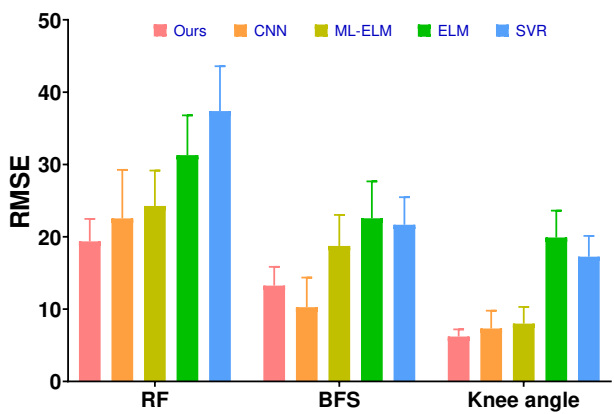}
\caption{Comparison results of intrasession scenario.}
\label{fig:9}
\end{figure}

The intrasession scenario is also considered to validate the robustness of the proposed framework. For each subject, the data with different walking speeds are fused into a whole dataset, where 80\% for training and the rest 20\% for testing.
Fig.~\ref{fig:9} depicts the corresponding experimental results, in which the proposed framework outperforms most baseline methods. According to Fig.~\ref{fig:9}, the proposed framework is not affected by the walking speeds, but the predicted results of some baseline methods are degraded. For instance, the predicted performance of SVR for muscle force of RF prediction illustrated in Fig.~\ref{fig:8}(a) is better than that of ELM, but its predicted performance becomes worse in the intrasession scenario due to the walking speeds.

\section{Discussions and Future Directions}
\label{sec:discussion}
In this section, we first discuss the effects of training dataset sizes on the prediction performance of the proposed framework and CNN. The flexibility of the proposed framework and the essential advantages of of physics-informed deep learning in musculoskeletal modelling are then presented, respectively. Finally, the limitations of this work and future directions are considered.

\subsection{Effects of Training Dataset Sizes on Performance}
\begin{figure}
\centering
\includegraphics[width=1\linewidth]{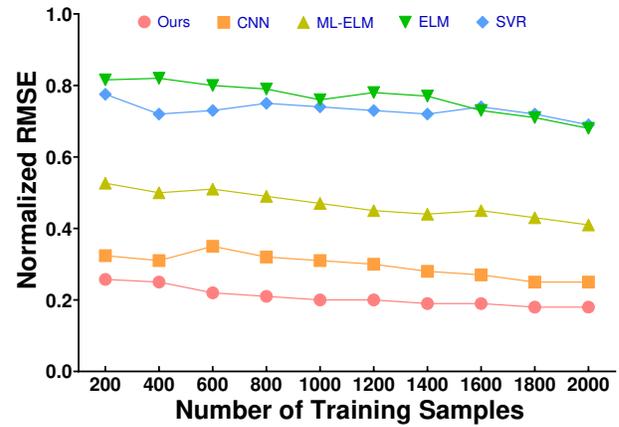}
\caption{Effects of the size of the training dataset on prediction performance of the wrist joint. The proposed framework achieves a faster convergence speed with simpler neural network architecture.}
\label{fig:11}
\end{figure}

To evaluate the effects of the training dataset sizes on the prediction performance, we illustrate the normalised RMSEs of wrist joint case of the proposed framework and baseline methods under different training data sizes in Fig.~\ref{fig:11}. The normalised RMSEs of both the proposed framework and baseline methods becomes low with the increase of training dataset sizes, but the proposed framework, on the other side, could achieve lower normalised RMSEs with fewer training samples.
Experimental results indicate the proposed framework is less sensitive to the training dataset size. Embedding physics-based domain knowledge as penalisation/regularisation term in the loss function of CNN yields faster convergence speed and reduces requirements in training data for a given performance.

\subsection{Flexibility of the Proposed Framework}
\begin{figure}
\centering
\includegraphics[width=1\linewidth]{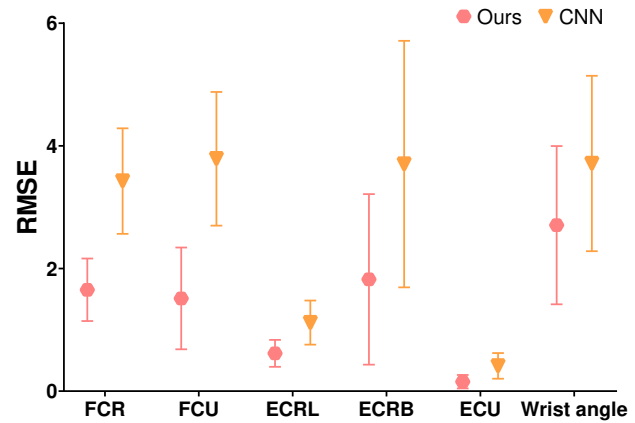}
\caption{Comparison results between the proposed framework and CNN under the same neural network architecture. The proposed framework achieves better predicted results without considering the effects of neural network architecture.}
\label{fig:12}
\end{figure}

The proposed framework is a generic paradigm for incorporating mechanistic musculoskeletal constraints. In this paper, we utilise a specific case, i.e., joint angle and muscle force prediction, as an exemplar to demonstrate the feasibility of this approach. Here, a CNN with simpler architecture is used as the data-driven modelling technique to extract feature maps from kinematics to muscle forces and EMG to motion, respectively. To further evaluate the proposed framework, we illustrate the comparison results of the wrist joint case of the proposed framework and CNN under the same neural network architecture, i.e., both CNN in the proposed framework and pure CNN have three convolutional blocks, three fully connected blocks, and one regression block, in Fig.~\ref{fig:12}. According to Fig.~\ref{fig:12}, the proposed framework could achieve better predicion performance than CNN, further indicating the effectiveness of the physics law in the proposed framework.

Additionally, all the components in the proposed framework can be withdrawn or adjusted, and new components could be incorporated into this framework depending on application demands. For example, in addition to CNN, the deep neural network in the data-driven component be replaced by CNN+LSTM when we want to extract spatial and temporal representations from kinematic measurements~\cite{bao2020cnn}, or generative adversarial network (GAN) when domain-independent features are required~\cite{creswell2018generative}. Additionally, few-shot learning should be considered when only a few training data are available~\cite{wang2020generalizing}, or federated learning is preferred when addressing data privacy issues~\cite{li2020federated}.
For the physics-based component, we can also incorporate the Hill muscle model into the proposed framework,  the predicted force should be equal to the Hill muscle model prediction, which can be used as the second physics law in the network.  In addition, more soft constraints can also be imposed, i.e., kinetic equations, activation dynamics, and contraction dynamics, to enhance the deep neural network performance.

\subsection{Understanding Physics-informed Data-driven Methods for Musculoskeletal Modelling}
Machine/deep learning methods have been utilised for musculoskeletal modelling, because they are conceptually intuitive simple and fast to implement~\cite{rane2019deep,dao2019deep,saxby2020machine}. However, while they may fit kinematic measurements very well in the training stage, their predictions may not satisfy the physics associated with musculoskeletal biomechanics. This can lead to poor robustness and generalisation. Data-driven methods provide limited interpretability from kinematic measurements. In contrast, physics-informed deep learning seamlessly integrates physics-based domain knowledge into deep learning and hence informative constraints for performance enhancement~\cite{wang2022and,jagtap2020conservative}. In physics-informed deep learning, we incorporate domain knowledge that from kinematic measurements and the physical understanding of the neuromusculoskeletal coupling. In this manner, machine/deep learning methods are interpretable, reflecting physical or physiological mechanisms, and more robust and generalise better even with limited training data.

\subsection{Limitations and Future Directions}
The proposed framework provides a generic paradigm for musculoskeletal modelling with the soft constraints imposed by physics-based domain knowledge.
Future work will develop the knowledge embedding deep neural networks via integrating more physics constraints. In addition, the proposed framework was only evaluated by simple one Degree of Freedom (DoF) movement, i.e., wrist and knee joint flexion and extension.  The reliability and accuracy of the proposed framework will also be evaluated in more complex scenarios involving multiple DoF movements.

Personalised musculoskeletal models are useful and helpful to assist and inform medical procedures, design subject-specific robotic devices, develop human-machine interfaces, etc.
However, creating the personalised musculoskeletal model is a challenging task.
This is attributed to the fact that accurate interpretations of musculoskeletal geometry for each specific patient require time-consuming and costly work~\cite{saxby2020machine}.
Different from existing methods, the proposed framework may significantly reduce the efforts of manual intervention. Such personalised information can be embedded into the framework to accelerate model building in a semi-automated manner.

\section{Conclusion}
\label{sec:conclusion}
This paper develops a knowledge embedding data-driven framework, which seamlessly integrates the physics-based domain knowledge into the data-driven model, for musculoskeletal modelling. Specifically, the physics-based domain knowledge is utilised as soft constraints to penalise/regularise the loss function of the deep neural network to enhance the robustness and generalisation performance, and computational demands in model building are significantly reduced. Comprehensive experiments on two groups of data for muscle forces and joint angles prediction indicate the feasibility of the proposed framework. We envision that the proposed framework is a general methodology for both muscle forces prediction and other applications in the musculoskeletal modelling field, which may reduce the gaps between laboratory prototypes and clinical applications.

\IEEEtriggeratref{30}

\bibliographystyle{IEEEtran}
\bibliography{IEEEabrv,reference}


%

%





\end{document}